\documentclass[aps,prb,a4paper,preprint,showpacs]{revtex4}
\usepackage{amsmath}
\usepackage{amsfonts}
\usepackage{amssymb}
\usepackage{slashbox}
\usepackage{graphicx}
\usepackage{amsbsy}

\begin{document}
\title{A scattering matrix approach to quantum pumping: Beyond the small-ac-driving-amplitude limit}
\author{Rui Zhu\renewcommand{\thefootnote}{*} \footnote{Corresponding author. Electronic address:
rzhu@scut.edu.cn}   }
\address{Department of Physics, South China University of Technology,
Guangzhou 510641, People's Republic of China }

\begin{abstract}

In the adiabatic and weak-modulation quantum pump, net electron flow
is driven from one reservoir to the other by absorbing or emitting
an energy quantum $\hbar \omega $ from or to the reservoirs. In our
approach, high-order dependence of the scattering matrix on the time
is considered. Non-sinusoidal behavior of strong pumping is
revealed. The relation between the pumped current and the ac driving
amplitude varies from power of $2$, $1$ to $1/2$ when stronger
modulation is exerted. Open experimental observation can be
interpreted by multi-energy-quantum-related processes.

\end{abstract}


\maketitle

\narrowtext

\section{Introduction}

Generally speaking, the transport of matter from low potential to
high potential excited by absorbing energy from the environment can
be described as a pump process. The driving mechanics of classic
pumps is straightforward and well understood\cite{Ref30}. The
concept of a quantum pump is initiated several decades
ago\cite{Ref1} with its mechanism involving coherent tunneling and
quantum interference. Research on quantum pumping has attracted
heated interest since its experimental realization in an open
quantum dot\cite{Ref2, Ref3, Ref4, Ref5, Ref6, Ref7, Ref8, Ref9,
Ref10, Ref11, Ref12, Ref13, Ref14, Ref15, Ref16, Ref17, Ref18,
Ref19, Ref20, Ref21, Ref22, Ref23, Ref24, Ref25, Ref26, Ref27,
Ref28, Ref29}.

The current and noise properties in various quantum pump structures
and devices were investigated such as the magnetic-barrier-modulated
two dimensional electron gas\cite{Ref5}, mesoscopic one-dimensional
wire\cite{Ref7, Ref23}, quantum-dot structures\cite{Ref6, Ref12,
Ref13, Ref29, Ref30}, mesoscopic rings with Aharonov-Casher and
Aharonov-Bohm effect\cite{Ref8}, magnetic tunnel
junctions\cite{Ref11}, chains of tunnel-coupled metallic
islands\cite{Ref26}, the nanoscale helical wire\cite{Ref27},the
Tomonaga-Luttinger liquid\cite{Ref25}, and garphene-based
devices\cite{Ref21, Ref22}. Theory also predicts that charge can be
pumped by oscillating one parameter in particular quantum
configurations\cite{Ref24}. A recent experiment\cite{Ref28} based on
two parallel quantized charge pumps offers a way forward to the
potential application of quantum pumping in quantum information
processing, the generation of single photons in pairs and bunches,
neural networking, and the development of a quantum standard for
electrical current. Correspondingly, theoretical techniques have
been put forward for the treatment of the quantum pumps\cite{Ref3,
Ref4, Ref19, Ref23, Ref26}. One of the most prominent is the
scattering approach proposed by Brouwer who presented
 a formula that relates the pumped current to the
parametric derivatives of the scattering matrix of the system.
Driven by adiabatic and weak modulation (the ac driving amplitude is
small compared to the static potential), the pumped current was
found to vary in a sinusoidal manner as a function of the phase
difference between the two oscillating potentials. It increases
linearly with the frequency in line with experimental finding.

Although the quantum pump has been extensively discussed in
literature, little attention was paid to experimentally observed
deviation from the weak-pumping theory with only the first-order
parametric derivative of the scattering matrix considered. We
improved the scattering approach by expanding the scattering matrix
to higher orders of the time and modulation amplitude, which enables
us to go further in investigation of the problem.

\section{Theoretical formulation}

We start with the scattering matrix approach detailed by Moskalets
\textit{et al}. \cite{Ref4} to describe the response of a mesoscopic
phase-coherent sample to two slowly oscillating (with a frequency
$\omega$) external real parameters $X_{j}(t)$ (gate potential,
magnetic flux, etc.),
\begin{equation}
\begin{array}{*{20}c}
   {X_j \left( t \right) = X_{0,j}  + X_{\omega ,j} e^{i\left( {\omega t - \varphi _j } \right)}  + X_{\omega ,j} e^{ - i\left( {\omega t - \varphi _j } \right)} ,} & {j = 1,2.}  \\
\end{array}
\end{equation}
$X_{0,j}$ and $X_{\omega ,j}$ measure the static magnitude and ac
driving amplitude of the two parameters, respectively. The phase
difference between the two drivers is defined as $\phi  = \varphi _1
- \varphi _2 $. The mesoscopic conductor is connected to two
reservoirs at zero bias. The scattering matrix $\hat s$ being a
function of parameters $X_{j}(t)$ depends on time.

It is assumed that the external parameter changes sufficiently
slowly to validate an ``instant scattering" description. To
investigate the deviation from the small amplitude $X_{\omega ,j} $
limit, we expand the scattering matrix $\hat s (t)$ into Taylor
series of $X_{j} (t)$ to second order at $X_{0,j}$ with the terms
linear and quadratic of $X_{\omega ,j} $ present in the expansion,
\begin{equation}
\hat s\left( t \right) \approx \hat s_{0}\left( {X_{0,j} } \right) +
\hat s_{ - \omega } e^{i\omega t}  + \hat s_{ + \omega } e^{ -
i\omega t}  + \hat s_2  + \hat s_{ - 2\omega } e^{2i\omega t}  +
\hat s_{ + 2\omega } e^{ - 2i\omega t} ,
\end{equation}
with
\begin{equation}
\left\{ \begin{array}{l}
 \hat s_{ \pm \omega }  = \sum\limits_{j = 1,2} {X_{\omega ,j} e^{ \pm i\varphi _j } {{\partial \hat s} \mathord{\left/
 {\vphantom {{\partial \hat s} {\partial X_j }}} \right.
 \kern-\nulldelimiterspace} {\partial X_j }}} , \\
 \hat s_2  = \sum\limits_{j = 1,2} {X_{\omega ,j}^2 {{\partial ^2 \hat s} \mathord{\left/
 {\vphantom {{\partial ^2 \hat s} {\partial X_j^2 }}} \right.
 \kern-\nulldelimiterspace} {\partial X_j^2 }}} , \\
 \hat s_{ \pm 2\omega }  = \frac{1}{2}\sum\limits_{j = 1,2} {X_{\omega ,j}^2 e^{ \pm 2i\varphi _j } {{\partial ^2 \hat s} \mathord{\left/
 {\vphantom {{\partial ^2 \hat s} {\partial X_j^2 }}} \right.
 \kern-\nulldelimiterspace} {\partial X_j^2 }}} . \\
 \end{array} \right.
\end{equation}
It can be seen from the equations that higher orders of the Fourier
spectra enter into the scattering matrix. As a result, both the
nearest and next nearest sidebands are taken into account, which
implies that a scattered electron can absorb or emit an energy
quantum of $\hbar \omega $ or $2 \hbar \omega $ before it leaves the
scattering region. In principle, third or higher orders in the
Taylor series can be obtained accordingly. However, the higher-order
parametric derivatives of the scatter matrix diminish dramatically
and approximate zero. Numerical calculation demonstrates that even
in relatively large amplitude modulation, their contribution is
negligible.

The pumped current depends on the values of the scattering matrix
within the energy interval of the order of $\max \left( {k_B T,2
\hbar \omega } \right)$ near the Fermi energy. In the
low-temperature limit ($T \to 0$), an energy interval of $2 \hbar
\omega $ is opened during the scattering process.

The mesoscopic scatterer is coupled to two reservoirs with the same
temperatures $T$ and electrochemical potentials $\mu$. Electrons
with the energy $E$ entering the scatterer are described by the
Fermi distribution function $f_{0} (E)$, which approximates a step
function at a low temperature. Due to the interaction with an
oscillating scatterer, an electron can absorb or emit energy quanta
that changes the distribution function. A single transverse channel
in one of the leads is considered. Applying the hypothesis of an
instant scattering, the scattering matrix connecting the incoming
and outgoing states can be written as
\begin{equation}
\hat b_\alpha  \left( t \right) = \sum\limits_\beta  {s_{\alpha
\beta } \left( t \right)\hat a_\beta  \left( t \right)}.
\end{equation}
Here $s_{\alpha \beta } $ is an element of the scattering matrix
$\hat s$; the time-dependent operator is $\hat a_\alpha  \left( t
\right) = \int {dE\hat a_\alpha  \left( E \right)e^{{{ - iEt}
\mathord{\left/
 {\vphantom {{ - iEt} \hbar }} \right.
 \kern-\nulldelimiterspace} \hbar }} } $,
and the energy-dependent operator ${\hat a_\alpha  \left( E
\right)}$ annihilates particles with total energy E incident from
the $\alpha$ lead into the scatter and obey the following
anticommutation relations
\begin{equation}
\left[ {\hat a_\alpha ^\dag  \left( E \right),\hat a_\beta  \left(
{E'} \right)} \right] = \delta _{\alpha \beta } \delta \left( {E -
E'} \right).
\end{equation}
Note that above expressions correspond to single- (transverse)
channel leads and spinless electrons. For the case of many-channel
leads each lead index ($\alpha $, $\beta$, etc.) includes a
transverse channel index and any repeating lead index implies
implicitly a summation over all the transverse channels in the lead.
Similarly an electron spin can be taken into account.

Using Eqs. (2) and (4) and after a Fourier transformation we obtain
\begin{equation}
\begin{array}{l}
 \hat b_\alpha  \left( E \right) = \sum\limits_\beta  {\left[ {\hat s_{0,\alpha \beta } \hat a_\beta  \left( E \right) + \hat s_{2,\alpha \beta } \hat a_\beta  \left( E \right) + \hat s_{ - \omega ,\alpha \beta } \hat a_\beta  \left( {E + \hbar \omega } \right)} \right.}  \\
 \left. { + \hat s_{ + \omega ,\alpha \beta } \hat a_\beta  \left( {E - \hbar \omega } \right) + \hat s_{ - 2\omega ,\alpha \beta } \hat a_\beta  \left( {E + 2\hbar \omega } \right) + \hat s_{ + 2\omega ,\alpha \beta } \hat a_\beta  \left( {E - 2\hbar \omega } \right)} \right]. \\
 \end{array}
\end{equation}
The distribution function for electrons leaving the scatterer
through the lead $\alpha$ is $f_\alpha ^{\left( {out} \right)}
\left( E \right) = \left\langle {\hat b_\alpha ^\dag  \left( E
\right)\hat b_\alpha  \left( E \right)} \right\rangle $, where
$\left\langle  \cdots  \right\rangle $ means quantum-mechanical
averaging. Substituting Eq. (6) we find
\begin{equation}
\begin{array}{l}
 f_\alpha ^{\left( {out} \right)} \left( E \right) = \sum\limits_\beta  {\left[ {\left| {\hat s_{0,\alpha \beta }  + \hat s_{2,\alpha \beta } } \right|^2 f_0 \left( E \right) + \left| {\hat s_{ - \omega ,\alpha \beta } } \right|^2 f_0 \left( {E + \hbar \omega } \right)} \right.}  \\
 \left. {\left| {\hat s_{ + \omega ,\alpha \beta } } \right|^2 f_0 \left( {E - \hbar \omega } \right) + \left| {\hat s_{ - 2\omega ,\alpha \beta } } \right|^2 f_0 \left( {E + 2\hbar \omega } \right) + \left| {\hat s_{ + 2\omega ,\alpha \beta } } \right|^2 f_0 \left( {E - 2\hbar \omega } \right)} \right]. \\
 \end{array}
\end{equation}
The distribution function for outgoing carriers is a nonequilibrium
distribution function generated by the nonstationary scatterer. The
Fourier amplitudes of the scattering matrix ${\left| {\hat s_{ -
\omega ,\alpha \beta } } \right|^2 }$ (${\left| {\hat s_{ + \omega
,\alpha \beta } } \right|^2 }$) is the probability for an electron
entering the scatterer through the lead $\beta$ and leaving the
scatterer through the lead $\alpha$ to emit (to absorb) an energy
quantum $\hbar \omega $ and ${\left| {\hat s_{ - 2\omega ,\alpha
\beta } } \right|^2 }$ (${\left| {\hat s_{ + 2\omega ,\alpha \beta }
} \right|^2 }$) is that of the energy quantum $2 \hbar \omega $
process. ${\left| {\hat s_{0,\alpha \beta }  + \hat s_{2,\alpha
\beta } } \right|^2 }$ is the probability for the same scattering
without the change of an energy with the second-order term $\hat
s_{2,\alpha \beta }$ much smaller than the zero-order term $\hat
s_{0,\alpha \beta }$ in weak-modulation limit ($X_{\omega ,j}  \ll
X_{0,j} $) and can be omitted therein.

Using the distribution functions $f_{0} (E)$ for incoming electrons
and $f_{\alpha} ^{out} (E)$ for outgoing electrons, the pumped
current measured at lead $\alpha$ reads
\begin{equation}
I_p  = \frac{e}{{2\pi \hbar }}\int_0^\infty  {\left\langle {\hat
b_\alpha ^\dag  \left( E \right)\hat b_\alpha  \left( E \right)}
\right\rangle  - \left\langle {\hat a_\alpha ^\dag  \left( E
\right)\hat a_\alpha  \left( E \right)} \right\rangle dE}.
\end{equation}
Substituting Eqs. (7) and (3) we get
\begin{equation}
\begin{array}{c}
 I_p  = \frac{{e\omega }}{{2\pi }}\sum\limits_{\beta ,j_1 ,j_2 } {X_{\omega ,j_1 } X_{\omega ,j_2 } \frac{{\partial s_{\alpha \beta } }}{{\partial X_{j_1 } }}\frac{{\partial s_{\alpha \beta }^* }}{{\partial X_{j_2 } }}2i\sin \left( {\varphi _{j_1 }  - \varphi _{j_2 } } \right)}  \\
  + \frac{{e\omega }}{{2\pi }}\sum\limits_{\beta ,j_1 ,j_2 } {X_{\omega ,j_1 }^2 X_{\omega ,j_2 }^2 \frac{{\partial ^2 s_{\alpha \beta } }}{{\partial X_{j_1 }^2 }}\frac{{\partial ^2 s_{\alpha \beta }^* }}{{\partial X_{j_2 }^2 }}i\sin \left[ {2\left( {\varphi _{j_1 }  - \varphi _{j_2 } } \right)} \right]} . \\
 \end{array}
\end{equation}
Quantum pumping properties beyond former theory based on first-order
parametric derivative of the scattering matrix are demonstrated in
Eq. (9). By taking higher orders of the Fourier spectrum of the
scattering matrix into consideration, double $\hbar \omega $ energy
quantum (or a $2 \hbar \omega $ energy quantum) emission
(absorption) processes coact with single $\hbar \omega $ quantum
processes. In the weak-modulation limit, the second term in the
right-hand side of Eq. (9) is small, which implies that double
$\hbar \omega $ quantum processes are weak and therefore not
observable. As the ac driving amplitude is enlarged, this term
increases markedly and contribution from double $\hbar \omega $
quantum processes takes effect. As a result, the dependence of the
pumped current on the phase difference between two driving
oscillations deviates from sinusoidal and changes from $\sin \phi $
to $\sin 2\phi $, which is observed in experiment\cite{Ref2}.
Moreover, the relation between the pumped current and the ac driving
amplitude $X_{\omega ,j} $ is reshaped. It is also seen that the
linear dependence of the pumped current on the oscillation frequency
holds for multi-quanta-related processes. In the next section,
numerical results of the pumped current in a
two-oscillating-potential-barrier modulated nanowire are presented
and comparison with experiment is given.

\section{Numerical results and interpretations}

We consider a nanowire modulated by two gate potential barriers with
equal width $L=20$ {\AA} separated by a $2L=40$ {\AA} width well
(see Fig. 1). The electrochemical potential of the two reservoirs
$\mu$ is set to be $60$ meV according to the resonant level within
the double-barrier structure. The two oscillating parameters in Eq.
(1) correspond to the two ac driven potential gates $X_{1,2} \left(
t \right) \to U_{1,2} \left( t \right)$ with all the other notations
correspond accordingly. We set the static magnitude of the two gate
potentials $U_{0,1}  = U_{0,2} = U_0  = 100$ meV and the ac driving
amplitude of the modulations equal $U_{\omega ,1}  = U_{\omega ,2} =
U_\omega $.

In Fig. 2, the dependence of the pumped current on the phase
difference between the two ac oscillations is presented. In
weak-modulation regime (namely $U_\omega  $ is small), sinusoidal
behavior dominates. Here, three relatively large $U_\omega  $ is
selected to reveal the deviation from the sinusoidal dependence.
(The magnitude of the pumped current mounts up in power-law relation
as a function of $U_\omega  $ as shown in Fig. 3. The sinusoidal
curve for small $U_\omega  $ would be flat and invisible in the same
coordinate range.) It can be seen from the figure that the $I_p$-$
\phi $ relation varies from sinusoidal ($\sin \phi $) to
double-sinusoidal ($\sin 2 \phi $) as the ac oscillation amplitude
is increased. The interpretation follows from Eq. (9). The single
$\hbar \omega $ quantum emission (absortion) processes feature a
sinusoidal behavior while the $2 \hbar \omega $ quantum emission
(absortion) processes feature a double-sinusoidal behavior when the
Fourier index is doubled. As $U_\omega  $ is increased, double
$\hbar \omega $ quantum processes gradually parallel and outweigh
the single $\hbar \omega $ quantum ones. It is also demonstrated
that when the single $\hbar \omega $ quantum processes have the
effect of $\sin \phi $ dependence, the double $\hbar \omega $
quantum processes induce a $-\sin 2 \phi $ contribution with a sign
flip, which can be understood from the sign change of the derivative
of the scattering matrix. The effect of three- and higher $\hbar
\omega $ quantum processes is small even for large $U_\omega $
comparable to $U_0$. The experimental observations\cite{Ref2} as a
deviation from the weak-modulation limit are revealed by our theory.

Experiment\cite{Ref2} also discovered that for weak pumping the
dependence of the pumped current on the pumping strength obeys a
power of 2 relation, as expected from the simple loop-area
argument\cite{Ref3}; for strong pumping, power of 1 and 1/2 relation
is observed beyond former theory. We presented in Fig. 3 the
numerical results based on our theory of the $I_p$-$U_{\omega}$
relation at a fixed $\phi $. To demonstrate its power-law
dependence, natural logarithm of
 the variables is applied. From Eq. (9), it can be seen that for
large ac driving amplitude $U_{\omega}$, contribution of double
$\hbar \omega $ quantum processes (formulated in the second term on
the right hand side of the equation) causes the $I_p$-$U_{\omega}$
relation to deviate from its weak-modulation limit, the latter of
which is $I_p \propto U_\omega ^2 $. For different phase difference
between the two ac drivers, the deviation is different. At $\phi
=\pi $ the pumped current is invariably zero regardless of the order
of approximation determined by time-reversal symmetry. At $\phi =\pi
/2$, $\sin 2\phi $ is exact zero, and no difference is incurred by
introducing higher order effect. If we shift the value of $\phi$ to
$0.49 \pi$, the abating effect of the double $\hbar \omega $ quantum
processes has the order of $U_\omega ^4 $ with the small
second-order parametric derivative of the scattering matrix
smoothing that effect a bit. Consequently, a power of $2 \to 1 \to
1/2$ relation is obtained and visualized by our curve fit, which is
analogous to experimental findings. For different values of $\phi$,
sharper abating and augmental effect occurs with analogous
mechanisms. It is possible that the experiment\cite{Ref2} was done
at the phase difference close to $\pi /2$ while trying to approach
maximal pumped current in the adiabatic and weak-pumping limit.

\section{Conclusions}

Based on the ac scattering approach, we go further to expand the
time-dependent scattering matrix to higher orders of the modulation
amplitude and the time. It is demonstrated in our theory that $2
\hbar \omega $ quantum emission (absorption) processes coact with
those of single $\hbar \omega $ quantum when we go beyond the
small-frequency and weak-modulation limit. Nonsinusoidal dependence
on the phase difference between two oscillating modulations is
incurred by higher order Fourier components. The pumped current
versus modulation amplitude relation has a power law of $2 \to 1 \to
1/2$ passage with the increase of the oscillating amplitude.
Numerical results for a two-ac-gate modulated nanowire interpret
experimental findings at large ac driving amplitudes.

\section{Acknowledgements}

The author would like to express sincere appreciation to Professor
Wenji Deng, Professor Jamal Berakdar, Professor Liliana Arrachea,
and Dr. Brian M. Walsh for valuable enlightenment to the topic from
discussions with them.

\clearpage

\begin{figure}[t]
\includegraphics{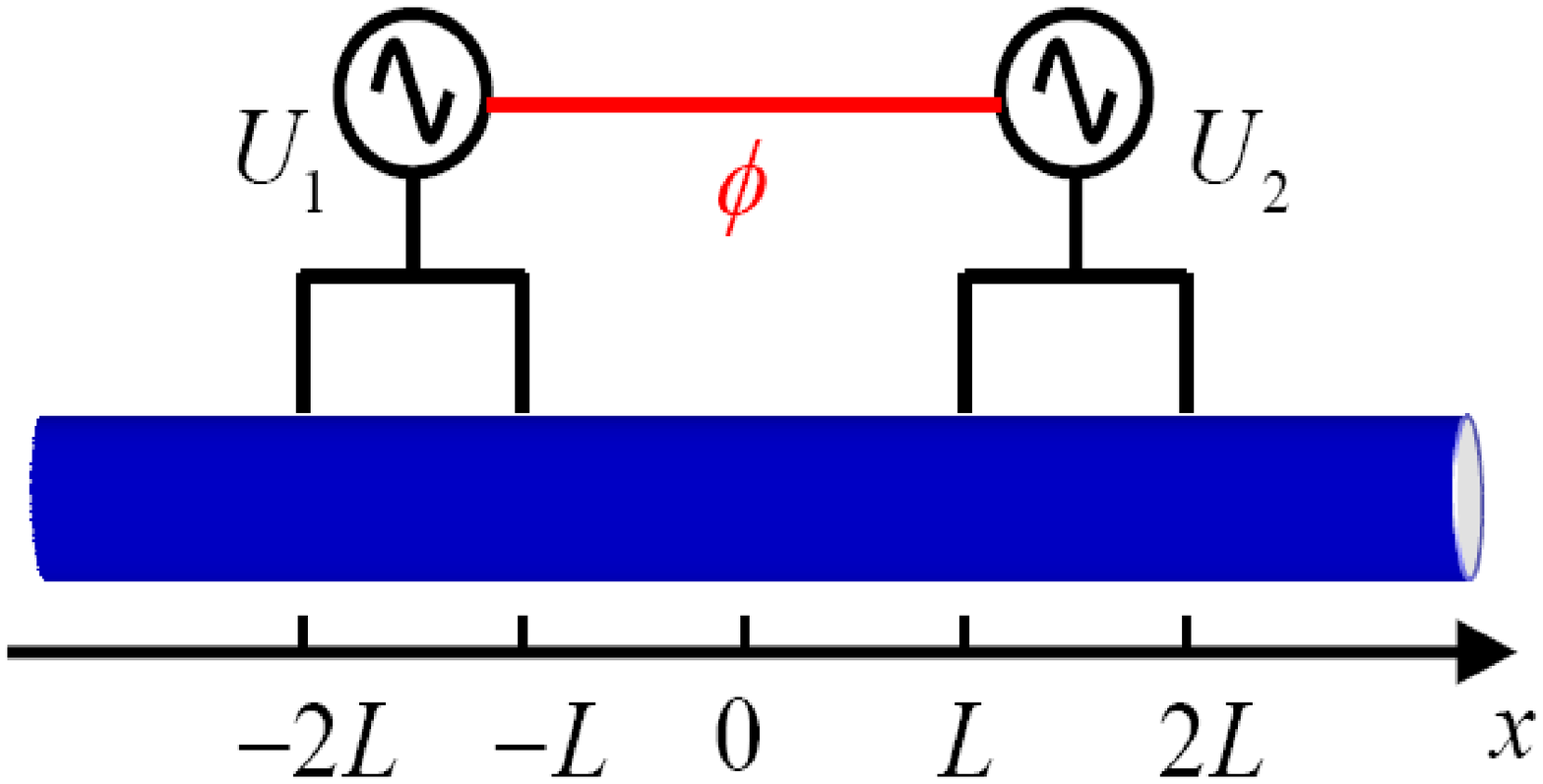}
\caption{Schematics of the quantum pump: a nanowire modulated by two
ac driven potential barriers.}
\end{figure}

\begin{figure}[h]
\includegraphics{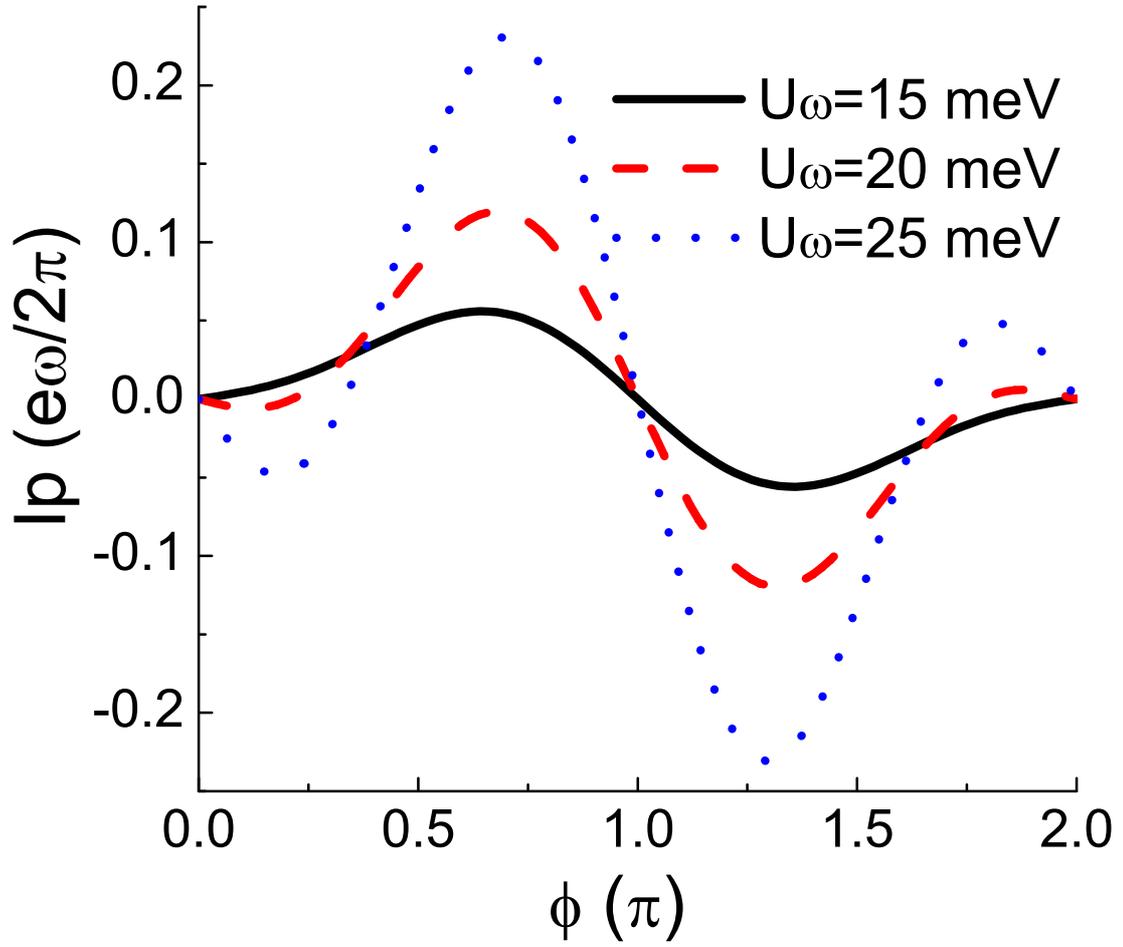}
 \caption{Pumped current as a function of the phase difference
 between the two modulations for different ac driving amplitudes.
 }
\end{figure}

\begin{figure}[h]
\includegraphics{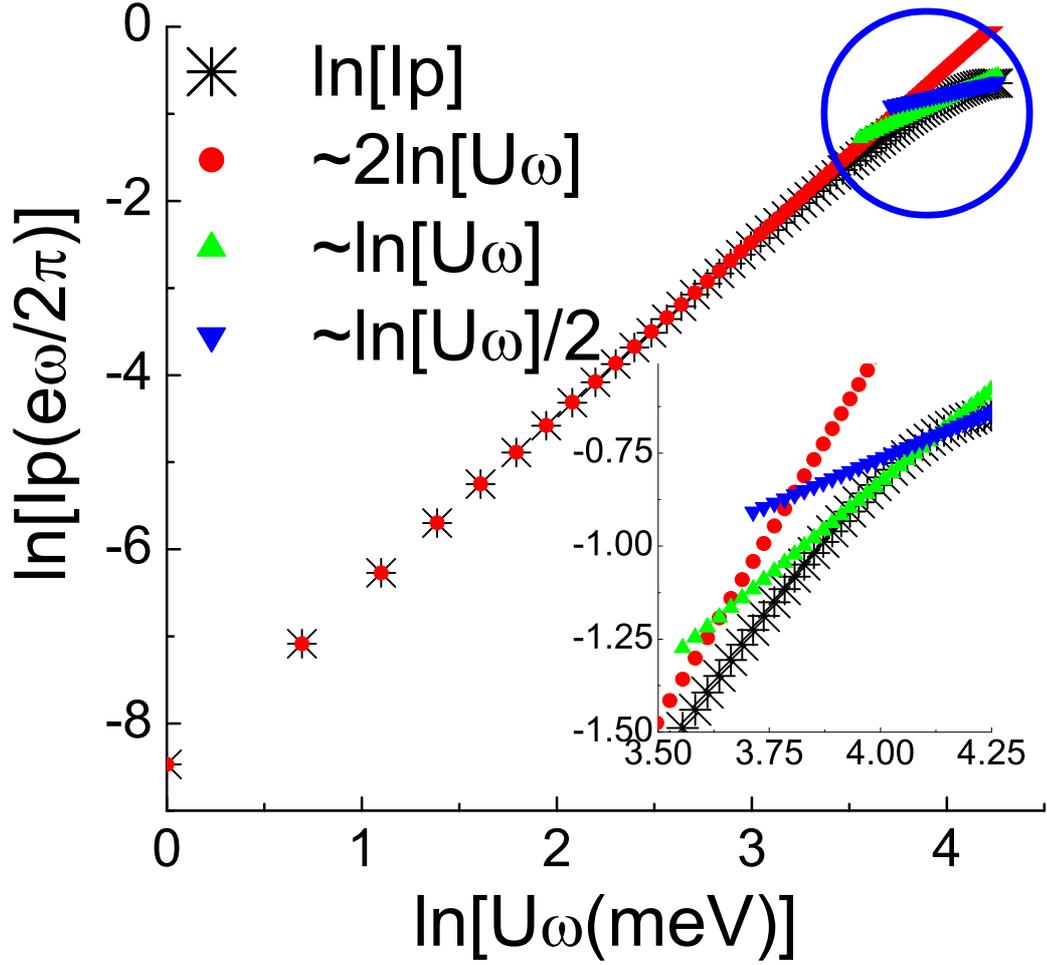}
 \caption{Pumped current as a function of the ac driving amplitude $U_\omega
 $ along with fits to $I_p  \propto U_\omega ^2 $ (red solid
 circle) below 35 meV,
$I_p  \propto U_\omega  $ (green upward triangle) below 41 meV, and
$I_p \propto U_\omega ^{1/2} $ above 41 meV (blue downward
triangle). To demonstrate its power-law dependence, natural
logarithm of
 the variables is applied.
 The phase difference between the two ac driver $\phi  = 0.49\pi $.
 Inset is the zoom-in of the circled region.}
\end{figure}

\end{document}